\documentclass[colorlinks,superbib,superscriptaddress,twocolumn,
nofootinbib,letterpaper,amsmath,amssymb]{revtex4}
\usepackage{amsfonts}
\usepackage{graphicx}
\usepackage{subfigure}
\usepackage{epsfig}
\usepackage{overpic}
\usepackage{amssymb,amsmath,bm,amsthm}
\usepackage{url}
\usepackage{color}
\usepackage{hyperref}

\theoremstyle{definition}
\newtheorem{defn}{Definition}

\begin{document}

\title{Estimation of the control parameter from symbolic sequences: Unimodal
maps with variable critical point}
\author{David Arroyo}
\email{david.arroyo@iec.csic.es}
\affiliation{Instituto de F\'{\i}sica Aplicada, Consejo Superior de Investigaciones Cient%
\'{\i}ficas, Serrano 144---28006 Madrid, Spain}
\author{Gonzalo Alvarez}
\affiliation{Instituto de F\'{\i}sica Aplicada, Consejo Superior de Investigaciones Cient%
\'{\i}ficas, Serrano 144---28006 Madrid, Spain}
\author{Jos\'{e} Mar\'{\i}a Amig\'{o}}
\affiliation{Centro de Investigaci\'on Operativa, Universidad Miguel Hern\'andez, Avda.
de la Universidad s/n, 03202 Elche, Spain}

\begin{abstract}
The work described in this paper can be interpreted as an application of the
order patterns of symbolic dynamics when dealing with unimodal maps.
Specifically, it is shown how Gray codes can be used to estimate the
probability distribution functions (PDFs) of the order patterns of
parametric unimodal maps. Furthermore, these PDFs depend on the value of the
parameter, what eventually provides a handle to estimate the parameter value
from symbolic sequences (in form of Gray codes), even when the critical
point depends on the parameter.
\end{abstract}

\keywords{Unimodal maps, symbolic dynamics, Gray codes, order patterns,
estimation of control parameter.}
\maketitle

\bfseries In this paper, the order patterns of unimodal maps are
studied. It is shown how to construct order patterns of unimodal
maps from their symbolic dynamics with respect to the partition of
the state space introduced by the critical point. Finally, it is
shown that for a subclass of parametric unimodal maps, the study of
those order patterns allows to estimate the parameter of the map
that has generated the symbolic sequence. \mdseries

\section{Introduction}

Sarkovskii's theorem shows that order and dynamics are intertwined
in one-dimensional intervals. It is therefore not surprising that
the study of the ordinal structure of deterministic time series
gives valuable information on the underlying dynamical system. This
work focuses on the reconstruction of the so-called order patterns
of certain unimodal maps, from \textquotedblleft
coarse-grained\textquotedblright\ orbits in form of 0-1 sequences: 0
if the corresponding iterate lies to the left of the critical point,
and 1 otherwise. Such binary sequences will be called Gray codes.
The relationship between the Gray codes of parametric unimodal maps
and the value of the parameter that controls a particular dynamic,
was shown in \cite{metropolis73,wang87,alvarez98}. Other important
tool for the understanding of one-dimensional dynamical systems is
the study of their order patterns \cite{amigo06}. Indeed, order
patterns allow to distinguish chaos from white noise, and can
provide useful information on the parameter or parameters
controlling the dynamic of chaotic systems. The main goal of this
paper is to estimate the control parameter of unimodal maps by means
of their order patterns alone, even when the exact values of their
orbits are not accessible but only the corresponding Gray codes.

The rest of the paper is organized as follows. First of all, the general
framework is set in Sect.~\ref{sec:scenario}. In Sect.~\ref%
{sec:orderPatterns}, the concept of order pattern is introduced, and its
dependence on the control parameter is analyzed for the logistic and the
skew tent maps. Sect.~\ref{sec:grayCodes} summarizes the theory on Gray
codes. How the order patterns of unimodal maps are obtained using Gray codes
is explained in Sect.~\ref{sec:grayandorder}; its application to control
parameter estimation is explained in Sect. VI. The results presented in this
paper are recapitulated in Sect.~\ref{sec:conclusions}, where some final
comments are also included.

\section{Scenario}

\label{sec:scenario} The work described in this paper focuses on a class of
\emph{unimodal maps}, hereafter denoted as $\mathcal{F}$. A map $%
f:I\rightarrow I$, where $I=[a,b]\subset \mathbb{R}$, $a<b$, belongs to the
class $\mathcal{F}$ if it satisfies the following conditions.

\begin{enumerate}
\item $f$ is continuous.

\item $f(a)=f(b)=a$.

\item $f$ reaches its maximum value $f_{\max }\leq b$ in the subinterval $%
[a_{m},b_{m}]\subset I$, $a_{m}\leq b_{m}$.

\item $f(f_{\max })<x_{c}$, where $x_{c}$ is the middle point of the
interval $[a_{m},b_{m}]$, i.e., $x_{c}=\frac{a_{m}+b_{m}}{2}$.

\item $f(x_{c})>x_{c}$.

\item $f$ is strictly increasing function on $[a,a_{m}]$ and strictly
decreasing on $[b_{m},b]$.
\end{enumerate}

The class $\mathcal{F}$ includes maps defined in a parametric way, say, $%
f_{\lambda }(x)=\varphi (\lambda ,x)$, where $x\in I=[a,b]$, $\lambda \in
J\subset \mathbb{R}$ is called the \emph{control parameter}, and $\varphi $
is a self-map of $I\times J$. Two different situations are considered in
this paper:

\begin{enumerate}
\item The control parameter determines the maximum value of the map. In this
case, the parametric function $f_{\lambda }$ is given by
\begin{equation}
f_{\lambda }(x)=\lambda F(x),
\end{equation}%
where $F\in \mathcal{F}$ and $F(x_{c})=F_{\max }$. The subclass of maps $%
f_{\lambda }\in \mathcal{F}$ complying with this description will be denoted
by $\mathcal{F}_{1}$.

\item The control parameter is the value of the critical point, i.e., $%
x_{c}=\lambda $. This leads to a new subclass of maps $\mathcal{F}_{2}$.
\end{enumerate}

\begin{figure}[!htb]
\centering
\begin{overpic}[scale=0.7]{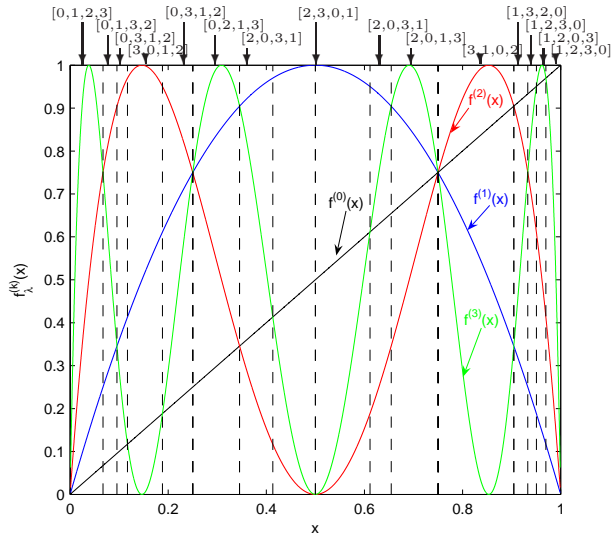}
        \put(10,85){\mbox{\tiny{[0,1,2,3]}}}
        \put(15,84){\vector(0,-1){6.6}}
        \put(17,83){\mbox{\tiny{[0,1,3,2]}}}
        \put(19,82){\vector(0,-1){4.6}}
        \put(20,81){\mbox{\tiny{[0,3,1,2]}}}
        \put(21,80){\vector(0,-1){2.6}}
        \put(22,79){\mbox{\tiny{[3,0,1,2]}}}
        \put(25,78.5){\vector(0,-1){1}}
        \put(28,85){\mbox{\tiny{[0,3,1,2]}}}
        \put(31,84){\vector(0,-1){6.6}}
        \put(34,83){\mbox{\tiny{[0,2,1,3]}}}
        \put(36,82){\vector(0,-1){4.6}}
        \put(40,81){\mbox{\tiny{[2,0,3,1]}}}
        \put(41,80){\vector(0,-1){2.6}}
        \put(49,85){\mbox{\tiny{[2,3,0,1]}}}
        \put(52,84){\vector(0,-1){6.6}}
        \put(60,83){\mbox{\tiny{[2,0,3,1]}}}
        \put(62,82){\vector(0,-1){4.6}}
        \put(66,81){\mbox{\tiny{[2,0,1,3]}}}
        \put(67,80){\vector(0,-1){2.6}}
        \put(75,79){\mbox{\tiny{[3,1,0,2]}}}
        \put(78,78.5){\vector(0,-1){1}}
        \put(82,85){\mbox{\tiny{[1,3,2,0]}}}
        \put(84,84){\vector(0,-1){6.6}}
        \put(85,83){\mbox{\tiny{[1,2,3,0]}}}
        \put(86,82){\vector(0,-1){4.6}}
        \put(87,81){\mbox{\tiny{[1,2,0,3]}}}
        \put(88,80){\vector(0,-1){2.6}}
        \put(89,79){\mbox{\tiny{[1,2,3,0]}}}
        \put(90,78.5){\vector(0,-1){1}}
        \end{overpic}
\caption{$f^{(k)}_\lambda(x)$ for $k=0,1,2,3$ and the corresponding
order patterns of length 4 for the logistic map when
$\protect\lambda=4$.} \label{figure:logistic_lambda_4}
\end{figure}

\section{Order patterns}

\label{sec:orderPatterns} Given a closed interval $I\subset \mathbb{R}$ and
a map $f:I\rightarrow I$ , the \emph{orbit} of (the initial condition) $x\in
I$ is defined as the set $\mathcal{O}_{f}(x)=\left\{ f^{n}(x):n\in \mathbb{N}%
_{0}\right\} $, where $\mathbb{N}_{0}=\{0\}\cup \mathbb{N}=\{0,1,...\}$, $%
f^{0}(x)=x$ and $f^{n}(x)=f\left( f^{n-1}(x)\right) $. Orbits are used to
define \emph{order} $L$-\emph{patterns }(or order patterns \emph{of length $L
$}), which are permutations of the elements $\{0,1,...,L-1\}$, $L\geq 2$. We
write $\pi =\left[ \pi _{0},\pi _{1},\ldots ,\pi _{L-1}\right] $ for the
permutation $0\mapsto \pi _{0},...,L-1\mapsto \pi _{L-1}$.

\begin{defn}[Order pattern]
The point $x\in I$ is said to define (or realize) the order $L$-pattern $\pi
=\pi (x)=\left[ \pi _{0},\pi _{1},\ldots ,\pi _{L-1}\right] $ if
\begin{equation}
f^{\pi _{0}}(x)<f^{\pi _{1}}(x)<\ldots <f^{\pi _{L-1}}(x).
\end{equation}%
Alternatively, $x$ is said to be of type $\pi $. The set of all possible
order patterns of length $L$ is denoted by $\mathcal{S}_{L}$.
\end{defn}

For further reference, it is convenient to assign an integer number to each
order pattern. This can be made, for instance, by means of the
Trotter-Johnson algorithm \cite{kreher:book}. The order patterns of length $4
$ along with their \textquotedblleft ordering numbers\textquotedblright ,
are shown in Table~\ref{table:orderPattern}.

\begin{table*}[tbh]
\centering
\begin{tabular}{|c|c||c|c||c|c||c|c|}
\hline
\# & Order pattern & \# & Order pattern & \# & Order pattern & \# & Order
pattern \\ \hline
0 & [0, 1, 2, 3] & 1 & [0, 1, 3, 2] & 2 & [0, 3, 1, 2] & 3 & [3, 0, 1, 2] \\
4 & [3, 0, 2, 1] & 5 & [0, 3, 2, 1] & 6 & [0, 2, 3, 1] & 7 & [0, 2, 1, 3] \\
8 & [2, 0, 1, 3] & 9 & [2, 0, 3, 1] & 10 & [2, 3, 0, 1] & 11 & [3, 2, 0, 1]
\\
12 & [3, 2, 1, 0] & 13 & [2, 3, 1, 0] & 14 & [2, 1, 3, 0] & 15 & [2, 1, 0, 3]
\\
16 & [1, 2, 0, 3] & 17 & [1, 2, 3, 0] & 18 & [1, 3, 2, 0] & 19 & [3, 1, 2, 0]
\\
20 & [3, 1, 0, 2] & 21 & [1, 3, 0, 2] & 22 & [1, 0, 3, 2] & 23 & [1, 0, 2, 3]
\\ \hline
\end{tabular}%
\caption{Order patterns of length four.}
\label{table:orderPattern}
\end{table*}
As emphasized in \cite{amigo08}, there always exist order $L$-patterns with
sufficiently large $L$ that are not realized in any orbit of $f\in \mathcal{F%
}$. These order patterns are called \emph{forbidden patterns}, whereas the
rest of order patterns are called \emph{allowed patterns}. In general, if $%
f_{\lambda }$ is a family of self-maps of the closed interval
$I\subset \mathbb{R}$ parameterized by $\lambda \in J\subset
\mathbb{R}$ (as it occurs for $f_{\lambda }\in
\mathcal{F}_{1},\mathcal{F}_{2}$), and the set $P_{\pi } $ is
defined as
\begin{equation}
P_{\pi }=\left\{ x\in I:x\text{ is of type }\pi \right\} ,
\label{eq:intervalPattern}
\end{equation}%
where $\pi \in \mathcal{S}_{L}$, then $P_{\pi }$ depends on $f_{\lambda }$
and, consequently, on $\lambda $. According to the \emph{ergodic theorem}
\cite[p. 34]{Walters}, if $f_{\lambda }$ is ergodic with respect to the
invariant measure $\mu $, then the orbit of $x\in I$ visits the set $P_{\pi }
$ with relative frequency $\mu \left( P_{\pi }\right) $, for almost all $x$
with respect to $\mu $. As a result, it is possible to study the dependence of $%
P_{\pi }$ on $\lambda $ by counting and normalizing the occurrences of $\pi $
in sliding windows of width $L$ along $\mathcal{O}_{f_{\lambda }}(x)$, $x$
being a `typical' initial condition. In the following two subsections this
is done experimentally with the logistic map (as representative of $\mathcal{%
F}_{1}$) and with the skew tent map (as representative of $\mathcal{F}_{2}$%
). Since we are primarily interested in the relation between the
probabilities $\mu (P_{\pi })$ (or relative frequencies) of order patterns $%
\pi \in \mathcal{S}_{L}$ and the control parameter $\lambda $ of the map
considered, we will refer to it as the $\lambda $-distribution function (in
short:\ $\lambda $-DF) of $\pi $, since they are related to the probability
distribution functions (we fix $\pi $ instead of fixing $\lambda $).

\begin{figure}[!hbt]
\centering
\includegraphics{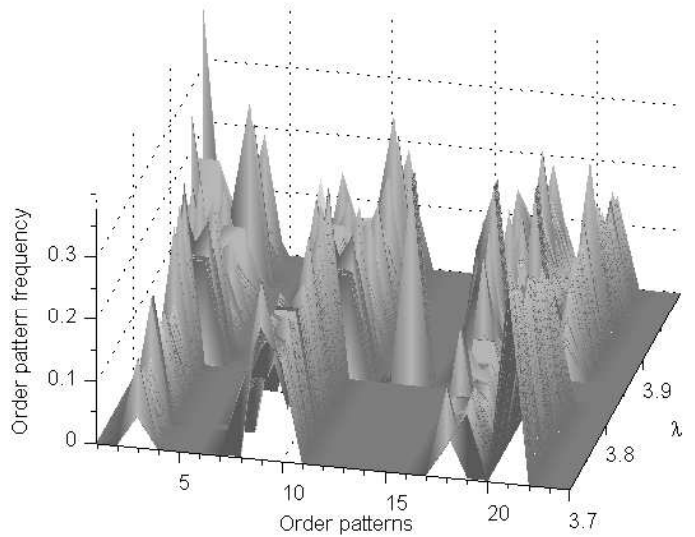}
\caption{Relative frequency of the order patterns realized by the logistic
map when $L=4$ and $\protect\lambda \in [3.7,4]$.}
\label{figure:logistic_orderPatternFreq}
\end{figure}

\subsection{Order patterns for the logistic map}

The logistic map, defined as
\begin{equation}
f_{\lambda }(x)=\lambda x(1-x),  \label{eq:logistic}
\end{equation}%
for $x\in \lbrack 0,1]$ and $\lambda \in \lbrack 1,4]$, belongs to $\mathcal{%
F}_{1}$. The logistic map with $\lambda =4$ was studied in \cite%
{amigo07,amigo08} from the ordinal point of view. In Fig.~\ref%
{figure:logistic_lambda_4} the allowed order $4$-patterns for the logistic
map with $\lambda =4$ are shown. For this value of the control parameter
there exist twelve allowed order patterns. However, the main goal of this
paper is to analyze the relationship between the control parameter of maps
in $\mathcal{F}_{1}$ or $\mathcal{F}_{2}$, and their order patterns, what
calls for the distributions of allowed patterns for different values of $%
\lambda $. Figure~\ref{figure:logistic_orderPatternFreq} depicts the
relative frequencies of each order $4$-pattern for $\lambda \in
\lbrack 3.7,4]$, the patterns being labeled as in
Table~\ref{table:orderPattern}. To be more specific, for every
$\lambda $, a sufficiently long orbit was generated, the occurrences
of the different order patterns were counted using a sliding window
of width 4, and finally the counts obtained were normalized by the
number of windows. These results are estimates of the probabilities
for the corresponding order patterns to occur. Let us point out
that, since the physical invariant measure of the logistic map is
only known for $\lambda =4$, numerical estimation of those
probabilities is the most we can hope for. More
importantly for us, we conclude from Fig. \ref%
{figure:logistic_orderPatternFreq} that it is very difficult to infer the
value of $\lambda \in \lbrack 3.7,4]$ from the $\lambda $-DF of order
patterns of length $4$.

\begin{figure*}[!htb]
\centerline{
\subfigure[$\lambda=0.3$]{\includegraphics[scale=0.9]{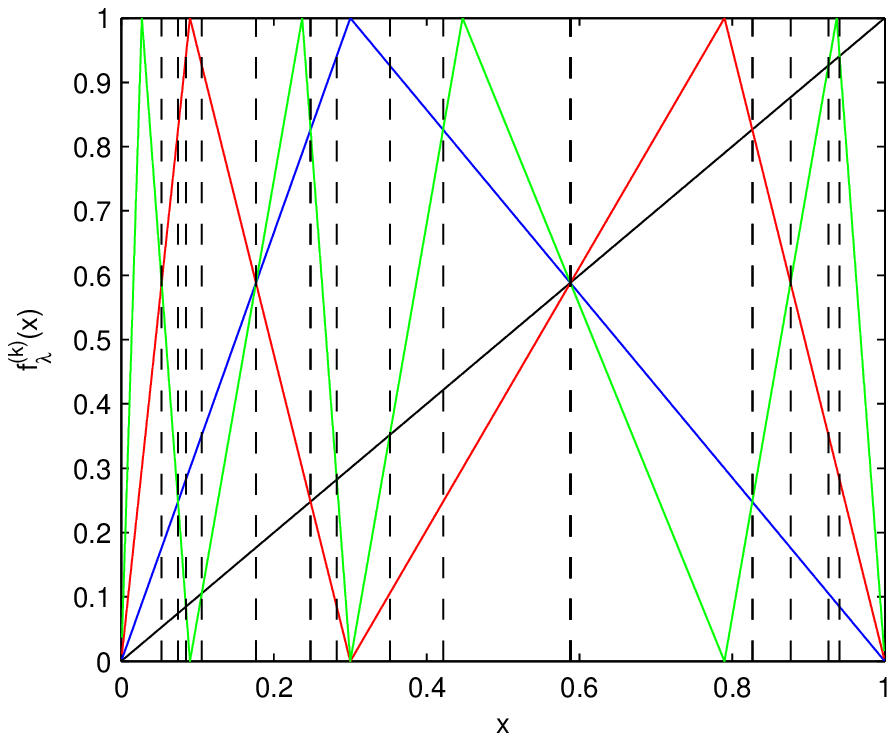}}
    \subfigure[$\lambda=0.7$]{\includegraphics[scale=0.9]{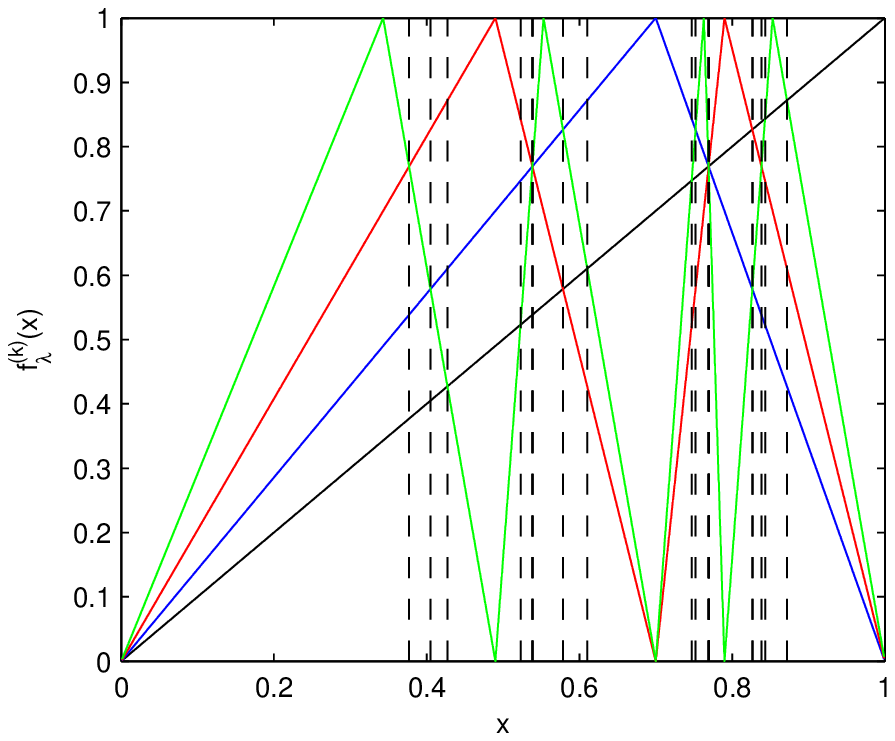}}
    }
\caption{The first four iterations of $f(x)$ and the corresponding order
patterns of length $4$ for the skew tent map, i.e., $f_\protect\protect%
\lambda^{(k)}(x)$ for $k=0,1,2,3$.}
\label{figure:skew_orderPatterns}
\end{figure*}

\subsection{Order patterns for the skew tent map}

\begin{figure}[!htb]
\centering
\includegraphics{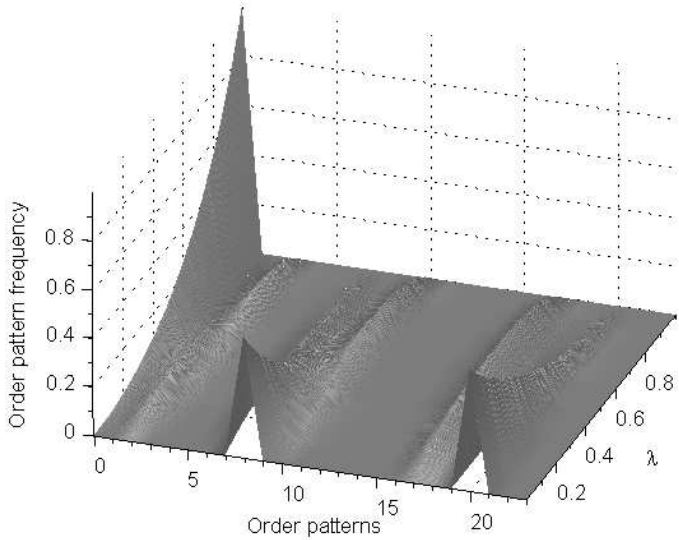}
\caption{Relative frequencies of the order patterns of length $L=4$ realized
by the skew tent map.}
\label{figure:skew_orderPatternFreq}
\end{figure}

\begin{figure}[!htb]
\centering
\includegraphics[width=10cm,height=9cm]{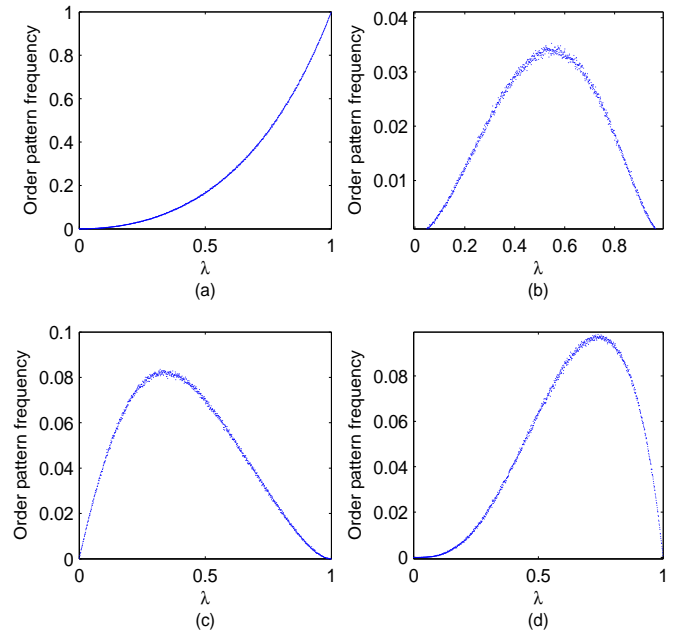}
\caption{Order pattern frequency for the skew tent map and $L=4$ (a) order
pattern $\#0$; (b) order pattern $\#1$; (c) order pattern $\#2$; (d) order
pattern $\#3$.}
\label{figure:skew_orderPatternFreq_detail}
\end{figure}

The skew tent map, given by
\begin{equation}
f_{\lambda }(x)=\left\{ {%
\begin{array}{lr}
x/\lambda , & \mbox{if $0 \leq x < \lambda$}, \\
(1-x)/(1-\lambda ), & \mbox{if $\lambda \leq x \leq 1$}, \\
\end{array}%
}\right.   \label{eq:skewTent}
\end{equation}%
for $x\in \lbrack 0,1]$ and $\lambda \in (0,1)$, belongs to the subclass $%
\mathcal{F}_{2}$, comprised of those maps of $\mathcal{F}$ parameterized by
the critical point. Furthermore, for the skew tent map $f_{\lambda }$, the
maximum value $f_{\lambda }(x_{c})=f_{\lambda }(\lambda )=1$ is independent
from $\lambda $ (see Fig.~\ref{figure:skew_orderPatterns}). Contrarily to
the logistic map, the skew tent map does possess a known ergodic invariant
measure for all $\lambda \in (0,1)$, namely, the Lebesgue measure on $[0,1]$%
. Hence, if $P_{\pi }$ is given by Eq.~\eqref{eq:intervalPattern} with $%
I=[0,1]$, the relative frequency of the order pattern $\pi $ in a typical
orbit of the skew tent map, coincides with the Lebesgue measure of $P_{\pi }$%
, which can be determined analytically. The easiest case corresponds
to the order pattern $\pi =[0,1,\ldots ,L-1]$, since then $P_{\pi }$
is an open interval whose left endpoint is $0$ and whose right
endpoint is the leftmost intersection between $f_{\lambda }^{(L-1)}$
and $f_{\lambda }^{(L-2)}$. The relative frequencies of the order
patterns of length 4, numbered according to Table I, are depicted in
Fig.~\ref{figure:skew_orderPatternFreq}. In particular, the length
of the interval $P_{[0,1,2,3]}=:(0,\phi _{4}(\lambda ))$ is
determined by the first intersection between $f_{\lambda }^{(2)}(x)$
and $f_{\lambda }^{(3)}(x) $:
\begin{equation}
\phi _{4}(\lambda )=\frac{\lambda ^{2}}{2-\lambda }.
\label{eq:orderPattern0skew}
\end{equation}%
Therefore, the $\lambda $-DF of $\pi =[0,1,2,3]$ (pattern $\#0$) is given by
$\phi _{4}(\lambda )$; see Fig.~\ref{figure:skew_orderPatternFreq_detail}(a)
for the graphical representation of $\phi _{4}(\lambda )$. The fact that the
function $\phi _{4}(\lambda )$ is bijective entails the possibility of
estimating $\lambda $ via the relative frequency of the order pattern $%
[0,1,2,3]$.

Up to this point it has been assumed that the orbits of the various maps
considered, were accessible. From a more practical point of view, it is also
relevant to know whether order patterns can be still determined using less
information about the orbits. This is the case, for instance, when dealing
with the symbolic dynamic associated to a generating partition of the state
space. In particular, the orbits of maps of $\mathcal{F}$ can be transformed
into binary sequences by the procedure described in \cite{metropolis73}. In
the next section it is explained how to build order patterns from those
binary sequences.

\section{Gray codes and unimodal maps}

\label{sec:grayCodes} Symbolic dynamics has been thoroughly studied in the
context of unimodal maps since the seminal contribution of Metropolis et al.
in \cite{metropolis73}. In \cite{alvarez98} Gray codes were used as a more
intuitive way of understanding and applying the ideas of \cite{metropolis73}%
. The connection between both approaches can be mathematically established
with the aid of results in \cite{metropolis73,BMS86,wang87}, as pointed out
in \cite{cusick99}. In this section, we address the ordinal structure of
Gray codes.

For a unimodal map $f$ defined on the interval $I=[a,b]$, any finite orbit $%
\{f^{n}(x):0\leq n\leq N-1\}$ can be transformed into a binary sequence $%
G_{N}(f,x)=g(f^{0}(x))\ g(f^{1}(x))\ \ldots g(f^{N-1}(x))$, where $g$ is the
step function
\begin{equation}
g(x)=\left\{
\begin{array}{cc}
0 & \mbox{ if }x<x_{c}, \\
1 & \mbox{ if }x\geq x_{c}.%
\end{array}%
\right.
\end{equation}%
As $x$ increases from the left endpoint $a$ to the right endpoint $b$, the
interval $I$ can be partitioned into $2^{N}$ subintervals $I_{j}^{(N)}$, $%
1\leq j\leq 2^{N}$, each subinterval containing those $x\in I$ whose orbits
have resulted into a given binary sequence $G_{N}(f,x)$. That is, (i) $%
I_{j}^{(N)}\cap I_{j}^{(N)}=\emptyset $ for $j\neq k$, (ii) $%
I=I_{1}^{(N)}\cup I_{2}^{(N)}\cdots \cup I_{2^{N}}^{(N)}$, and (iii) the
binary sequences $G_{N}(f,x)$ obtained for each $x\in I_{j}^{(N)}$ are the
same. Moreover, the sequences $G_{N}(f,x_{1})$ for $x_{1}\in I_{j}^{(N)}$
and $G_{N}(f,x_{2})$ for $x_{2}\in I_{j+1}^{(N)}$, $1\leq j\leq 2^{N}-1$,
differ only in one bit. Therefore, if we label the $2^{N}$ subintervals $%
I_{j}^{(N)}$ with the $2^{N}$ sequences $G_{N}(f,x)$, then the labels of
continuous subintervals will have only one bit flipped.

\begin{figure}[!htb]
\centering \epsfig{file=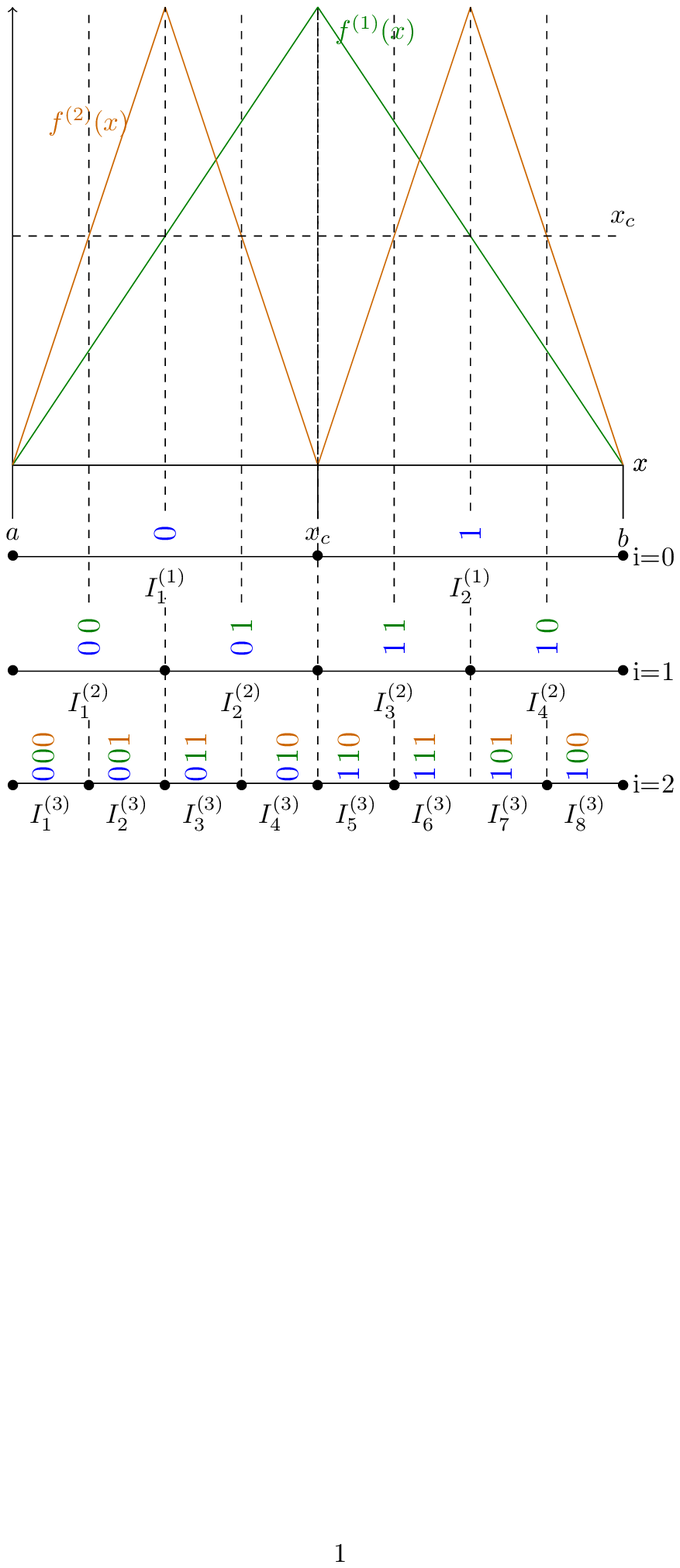} \caption{Symbolic
intervals for different iterations of the skew tent map for
$\protect\lambda=0.5$.} \label{figure:skewtentMapSymbolic}
\end{figure}

For the sake of illustration, let us consider the skew tent map with $%
\lambda =0.5$. In Fig.~\ref{figure:skewtentMapSymbolic}, the division of $%
I=[0,1]$ into the subintervals $I_{j}^{(N)}$, each labeled with the
corresponding binary sequence of length $N$, is shown for $N=1,2,3$. The
separation points of the subintervals $I_{j}^{(N)}$ are the solutions of the
equations
\begin{equation}
f_{1/2}^{n-1}(x)=\tfrac{1}{2},\;\;1\leq n\leq N.
\end{equation}

If, furthermore, $\mathcal{G}_{N}$ is the set of all binary sequences of
length $N$ produced by a map $f\in \mathcal{F}$, then it is possible to
endow $\mathcal{G}_{N}$ with a linear order as follows. Given $%
G_{N}(f,x_{1})\neq G_{N}(f,x_{2})$, let $i$ be the first index such that $%
g(f^{i}(x_{1}))\neq g(f^{i}(x_{2}))$. Depending on the value of $i$, we
distinguish three cases:

\begin{description}
\item[- ] If $i=0$ then $G_{N}(f,x_{1})<G_N(f,x_{2})$ if and only if $%
g(x_{1})<g(x_{2})$.

\item[- ] If $i>0$ and $G_{i}(f,x_{1})=G_{i}(f,x_{2})$ contains an even
number of $1$'s, then $G_{N}(f,x_{1})<G_{N}(f,x_{2})$ if and only if $%
g(f^{i}(x_{1}))<g(f^{i}(x_{2}))$.

\item[- ] If $i>0$ and $G_{i}(f,x_{1})=G_{i}(f,x_{2})$ contains an odd
number of $1$'s, then $G_{N}(f,x_{1})<G_{N}(f,x_{2})$ if and only if $%
g(f^{i}(x_{1}))>g(f^{i}(x_{2}))$.
\end{description}

Gray codes are well known in the context of communication theory.
The Gray codes of length $3$ are shown in
Table~\ref{table:Graycodes}. The main characteristic of the Gray
codes is that two consecutive codes differ in only one bit.
Moreover, the order of Gray codes is equivalent to the order in
$\mathcal{G}_{N}$ (check Table~\ref{table:Graycodes} for $N=3$). As
a consequence, any binary sequence $G_{N}(f,x)$ can be interpreted
as a Gray code of length $N$ \cite{alvarez98}, and will be called a
Gray code hereafter. Finally, the
order of the Gray codes derived from any unimodal map belonging to $\mathcal{%
F}$ is directly linked to the order in $\mathbb{R}$ of the points $x\in I$.
Indeed, it is proven in \cite[Lemma 4.1]{BMS86} that $%
G_{N}(f,x_{1})<G_{N}(f,x_{2})$ for some $N\geq 1$, implies $x_{1}<x_{2}$.
This is illustrated in Fig.~\ref{figure:skewtentMapSymbolic}.

\begin{table}
\centering
\begin{tabular}{|c|c|c|}
\hline
\textbf{Rank} & \textbf{Binary code} & \textbf{Gray code} \\ \hline
0 & 000 & 000 \\ \hline
1 & 001 & 001 \\ \hline
2 & 010 & 011 \\ \hline
3 & 011 & 010 \\ \hline
4 & 100 & 110 \\ \hline
5 & 101 & 111 \\ \hline
6 & 110 & 101 \\ \hline
7 & 111 & 100 \\ \hline
\end{tabular}%
\caption{Correspondence between Gray codes and binary codes for
three bits.} \label{table:Graycodes}
\end{table}

\begin{figure}[!htb]
\centering
\includegraphics[scale=.7]{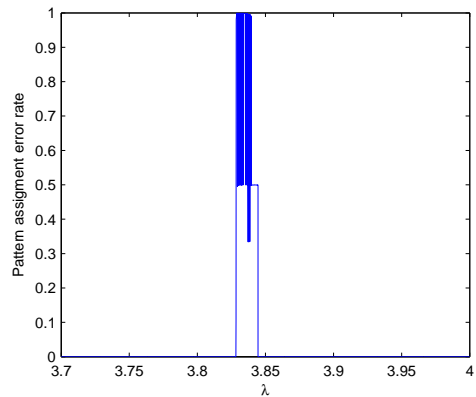}
\caption{Error rate for the pattern assignment based on Gray codes with
respect to the one based on the orbit of the logistic map. The length of
order patterns is $L=4$, the length of the considered Gray codes is $N=100$
and the number of samples is $10000$. The perfect estimation of the PDF of
the order patterns of the logistic map is possible for those values of $%
\protect\lambda$ leading to aperiodic binary sequences or to binary
sequences with period larger than $4$, i.e., the length of the considered
order patterns.}
\label{figure:errorLog}
\end{figure}

\section{Gray codes and order patterns for unimodal maps}

\label{sec:grayandorder} In this section the analysis focuses on the
parametric unimodal maps of the subclasses $\mathcal{F}_{1}$ or $\mathcal{F}%
_{2}$. In section \ref{sec:orderPatterns} we elaborated on the dependence of
the order patterns allowed for those maps with respect to the control
parameter. Specifically, we estimated the probabilities of order $4$%
-patterns by their relative frequencies in orbits of the logistic map (Fig. %
\ref{figure:logistic_orderPatternFreq}) and of the skewed tent map
(see Fig.~\ref{figure:skew_orderPatternFreq}) with different
parameter settings. Our next goal is to reproduce the same
dependencies not from the exact values of the orbit point
(\textquotedblleft sharp orbit\textquotedblright ), but from the
binary sequence built as explained in the previous section
(\textquotedblleft coarse-grained orbit\textquotedblright ). As
discussed in that section, the definition domain $I$ of $f\in
\mathcal{F}$ splits in $2^{N}$ subintervals when Gray
codes of length $N$ are considered. We show next that the order patterns of $%
f$ can also be obtained comparing Gray codes obtained from its orbits.

Let $G_{M}(f,x)=g_{0}g_{1}\ldots g_{M-1}$, $g_{i}\in \{0,1\}$, be the Gray
code of length $M$ of $x\in I$. Since the Gray codes, together with the
points $x\in I$, are linearly ordered and, moreover, their order relations
are equivalent (i.e., $G_{\infty }(f,x_{1})<G_{\infty }(f,x_{2})$ iff $%
x_{1}<x_{2}$), we can expect to obtain useful information about the
order patterns realized by the sharp orbit $\mathcal{O}_{f}(x)$ from
the order patterns realized by the coarse-grained orbit
$G_{M}(f,x)$, $M\geq 2$. The procedure is as follows.

\begin{enumerate}
\item Divide the Gray code of length $M$, $G_{M}(f,x)$, into $M-N+1$ Gray
codes of length $N<M$ using a sliding window of length $N$. Thus, the first
Gray code derived from $G_{M}(f,x)$ is $G^{0}=g_{0}g_{1}\ldots
g_{N-1}=G_{N}(f,x)$, the second Gray code is $G^{1}=g_{1}g_{2}\ldots
g_{N}=G_{N}(f,f(x))$, \ldots , and the $(M-N+1)$-th Gray code is $%
G^{M-N}=g_{M-N}g_{M-N+1}\ldots g_{M-1}=G_{N}(f,f^{M-N}(x))$.

\item For $i=0,1,...,M-N-L+1$, build groups of $L$ consecutive Gray codes $%
G^{i}G^{i+1}\ldots G^{i+L-1}$. The $i$-th group defines then the order $L$%
-pattern $\pi =\pi (i)=[\pi _{0},\pi _{1},\ldots ,\pi _{L-1}]$ if
\begin{equation*}
G^{i+\pi _{0}}<G^{i+\pi _{1}}<\ldots <G^{i+\pi _{L-1}}.
\end{equation*}
\end{enumerate}

The order patterns derived using Gray codes need not have, in
general, similar $\lambda $-DFs to those derived from the sharp
orbits. Indeed, order patterns defined by Gray codes of length $N$
are built upon the comparison of subintervals $I_{j}^{(N)}\subset I$
(see Sect.~\ref{sec:grayCodes}), rather than comparing points of
$I$. The width of the intervals $I_{j}^{(N)}$ decreases as the
length $N$ of the sliding window increases in such a way that when $%
N\rightarrow \infty $, each one of those intervals converges to a single
real number. As a result, the error in the calculation of the order patterns
from Gray codes is expected to reduce as $N$ increases. In the context of
finite-precision computation, the minimum value of $N$ necessary to get a
reliable approximation of the $\lambda $-DF of an order pattern is related
to the precision of the arithmetic used. Again, this quantization error
decreases as $N$ increases and, consequently, a large value of $N$ may be
necessary to assure a good approximation of the $\lambda $-DF.

\begin{figure}[!htb]
\centering
\includegraphics[width=8cm,height=6cm]{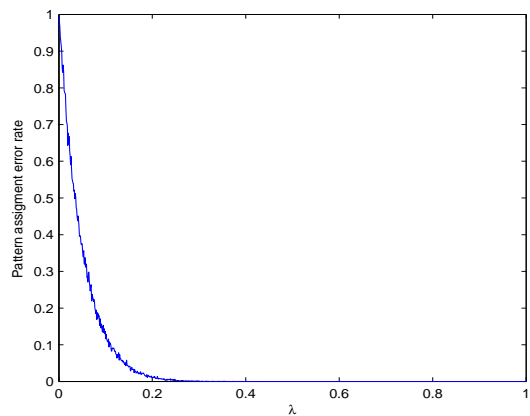}
\caption{Error rate for the pattern assignment based on Gray codes with
respect to the one based on the orbit of the skew tent map. The length of
order patterns is $L=4$, the length of the considered Gray codes is $N=100$
and the number of samples is $10000$. A value of the control parameter above
$0.2$ guarantees a perfect estimation of the PDF of the order patterns of
the skew tent map.}
\label{figure:errorSkew}
\end{figure}

\begin{figure}[!htb]
\centering
\includegraphics[scale=0.7]{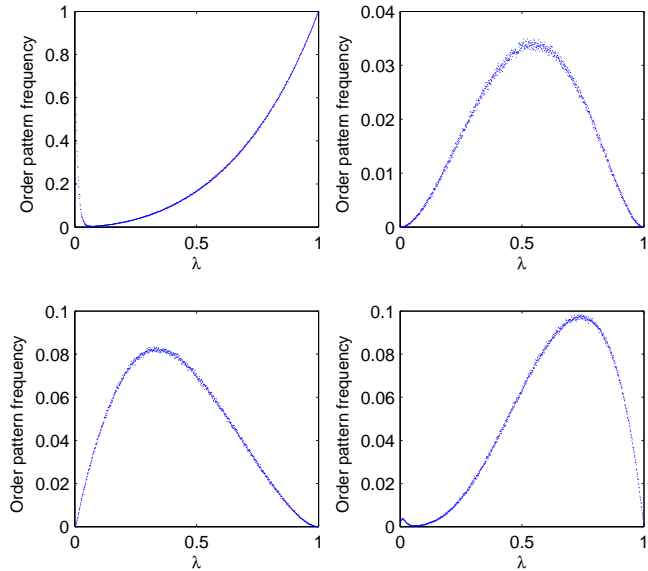}
\caption{Relative frequency of order patterns of the skew tent map using
Gray codes, when $L=4$, $N=100$ and the sequences are $10104$-bit long: (a)
order pattern $\#0$; (b) order pattern $\#1$; (c) order pattern $\#2$; (d)
order pattern $\#3$.}
\label{figure:skewOrderFrecGray}
\end{figure}

Another source of divergences between $\lambda $-DFs and their numerical
estimation via finite-length Gray codes maybe non-ergodicity or even poor
ergodicity. As a matter of fact, remember that the estimation of the
probability $\mu (P_{\pi })$ by the relative frequency of $\pi \in \mathcal{S%
}_{L}$ in finite orbits of a $\mu $-preserving map, hinges on the
ergodic theorem. If, furthermore, the convergence of relative
frequencies to probabilities in the orbits of an ergodic map with
respect to $\mu $, is very slow, a good estimation would require
exceedingly long sequences ---this is what we mean by
\textquotedblleft poor ergodicity\textquotedblright . These errors
are shown in Figs.~\ref{figure:errorLog} and ~\ref{figure:errorSkew}
for the logistic and the skew tent maps, respectively, with $\pi =[0,1,2,3]$%
, $M=10104$, and $N=100$. In the first case, the value of $\lambda $ lies
within the period-$3$ window of the logistic map. In the second case, poor
ergodicity is expected for values of $\lambda $ close to $0$ and $1$. The
asymmetry in the error distribution is due to the fact that for $\lambda
\simeq 1$, the tent map looks like the identity in most of $I=[0,1]$, hence $%
P_{[0,1,2,3]}$ covers most of $I$. This makes $[0,1,2,3]$ to be the most
frequent order $4$-pattern even when its frequency is calculated using Gray
codes. Comparison of Figs.~\ref{figure:skew_orderPatternFreq_detail} and ~%
\ref{figure:skewOrderFrecGray} illustrates the accuracy of the Gray
code-based method for the first four order $4$-patterns (see
Table~\ref{table:Graycodes}) of the skew tent map .

\section{Estimation of the control parameter for unimodal maps with critical
point depending on the control parameter}

\label{sec:parameterEstimation} The main characteristic of maps in $\mathcal{%
F}_{2}$ is that the control parameter $\lambda $ determines the
value of the critical point. Furthermore, from our discussion above,
we expect that the relation between the control parameter and the
allowed order patterns of the corresponding dynamics is specially
simple for the pattern $\pi =[0,1,...,L-1] $. Clearly, if the
$\lambda $-DF of this pattern is 1-to-1, then $\lambda $ can be
pinpointed from that distribution function; otherwise, the possible
values of $\lambda $ can be reduced to a few candidates, what can be
also acceptable in applications like cryptanalysis. In turn,
$\lambda $-DFs can be approximated via Gray codes, without previous
knowledge of the critical
point of the map. The bottom line is that the control parameter of a map in $%
\mathcal{F}_{2}$ can be estimated from their coarse-grained orbits (in form
of Gray codes). The specifics depend on the map.

As an example, consider the skew tent map again. For this map, the interval $%
P_{[0,1,\ldots ,L-1]}$, i.e., the set of points $x\in \lbrack 0,1]$ of type $%
[0,1,\ldots ,L-1]$, is determined by the leftmost intersection of the
iterates $f_{\lambda }^{L-2}$ and $f_{\lambda }^{L-1}$, where
\begin{equation}
f_{\lambda }^{n}(x)=\left\{
\begin{array}{ll}
x/\lambda ^{n}, & \text{ if }0\leq x\leq \lambda ^{n}, \\
(\lambda ^{n-1}-x)/\lambda ^{n-1}(1-\lambda ), & \text{ if }\lambda ^{n}\leq
x\leq \lambda ^{n-1}.%
\end{array}%
\right.
\end{equation}%
Hence $P_{[0,1,\ldots ,L-1]}=[0,\phi _{L}(\lambda )]$, with
\begin{equation}
\phi _{L}(\lambda )=\frac{\lambda ^{L-2}}{2-\lambda }.  \label{eq:upperBound}
\end{equation}%
Since this function is 1-to-1 in the interval $0\leq \lambda \leq 1$ for $%
L\geq 2$, with $\phi _{2}(0)=1/2$, $\phi _{L\geq 3}(0)=0$, and $\phi _{L\geq
2}(1)=1$, it allows to estimate $\lambda $ by estimating $\phi _{L}(\lambda
) $ ---the length of $P_{[0,1,\ldots ,L-1]}$. Now, from the equation
\begin{eqnarray}
\frac{d}{d\lambda }\phi _{L}(\lambda ) &=& \frac{\lambda
^{L-3}}{(2-\lambda )^{2}}[2(L-2)-(L-3)\lambda ]= \nonumber\\
&=&\left\{
\begin{array}{ll}
0, & \text{ if }\lambda =0, \\
L-1, & \text{ if }\lambda =1,%
\end{array}
\right.  \label{eq:deriv}
\end{eqnarray}%
it follows that $\phi _{L}(\lambda )$ is a $\cup $-convex function on $0\leq
\lambda \leq 1$ for $L\geq 2$, that converges to $0$ on $0\leq \lambda <1$
as $L\rightarrow \infty $. Therefore, the higher $L$ the worse $\phi
_{L}(\lambda )$ discriminates different values of $\lambda $. Consequently, $%
L=3,4$ are the best choices for a quality estimation of $\lambda $.

\begin{figure}[!htb]
\subfigure[]{\includegraphics[width=7cm,height=6cm]{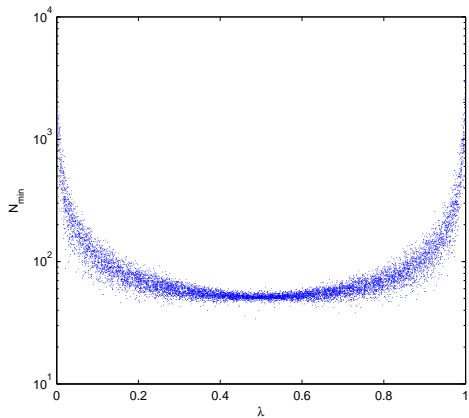}}
    \subfigure[]{\includegraphics[scale=.7]{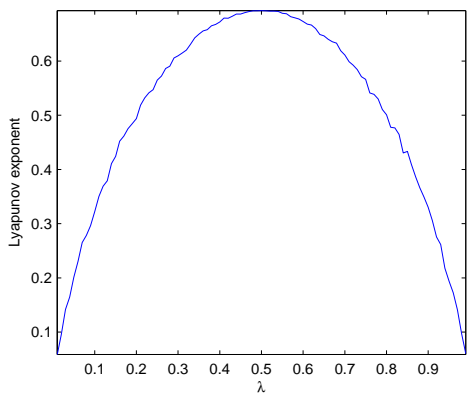}}
\caption{Dependence of the width of the sliding window with respect to the
rate of divergence:(a) Minimum width of the sliding window necessary for the
reconstruction of the PDF of the order patterns from the symbolic sequences
of the skew tent map; (b) Lyapunov exponent of the skew tent map.}
\label{figure:Nmin}
\end{figure}

On the other hand, the ergodicity of the skew tent map permits to
estimate the length of $P_{[0,1,\ldots ,L-1]}$ by estimating the
relative frequency of the $\pi =[0,1,\ldots ,L-1]$ in a typical
sharp orbit of the map ---or, as we intent, in a typical
coarse-grained orbit. In the latter case, the choice for the
parameter $N$, the width of the sliding window down the Gray codes
(Sect.~\ref{sec:grayandorder}), must be also analyzed. The minimum
value of $N$ to get a good reconstruction of the $\lambda $-DF of
the order patterns, $N_{min}$, depends on the precision of the
arithmetic used, but it also depends on the Lyapunov exponent of the
map. If floating point double-precision arithmetic is implemented,
then $N_{min}$ can be determined as function of $\lambda $ by
comparing pairs of symbolic sequences generated from the same
initial condition and control parameters $\lambda _{1}$ and $\lambda
_{2}$ such that $|\lambda _{2}-\lambda _{1}|$ equals the spacing of
floating point numbers. As it is shown in Fig.~\ref{figure:Nmin},
the value of $N_{min}$ increases with the Lyapunov exponent for the
skewed tent map.

Summing up, the estimation of the control parameter $\lambda \in (0,1)$ of
the skew tent map $f_{\lambda }$, Eq.~\eqref{eq:skewTent}, can be done by
counting and normalizing the occurrences of the order pattern $[0,1,...,L-1]$%
, ideally for $L=3$ or $4$, in a statistically significant sample of orbit
segments of $f_{\lambda }$. This follows from the following properties: (i) $%
f_{\lambda }$ is ergodic for all $\lambda $, and (ii) the $f_{\lambda }$%
-invariant measure of $P_{[0,1,...,L-1]}$ (in this case, the length of the
interval $P_{[0,1,...,L-1]}$) depends bijectively on $\lambda $. In a
practical context though, finite precision machines are used, and this
entails, in general, numerical degradation, this meaning that the computed
orbits, whether of chaotic or non-chaotic maps, depart from the real ones.
In the case of a very long orbit of a chaotic map, the deviation of the
numerical simulation (locally measured by the Lyapunov exponent of the map)
will be severe; in such cases, it is preferable to have many shorter orbits
instead. Even worse, all orbits computed with finite precision are
eventually periodic. This distortion of the dynamics, due to finite
numerical precision and dependence on initial conditions, implies the
general impossibility of obtaining orbits and invariant measures in an exact
way. As a matter of fact, all this carries over to symbolic dynamics.

To verify this issue in the case of coarse-grained orbits, a sample of Gray
codes of the skew tent map, each one with the same length but with a
different initial condition, was generated for every value of $\lambda $.
The underlying sharp orbits were computed with double precision floating
point arithmetic. From this sample of Gray codes, the corresponding $\lambda
$-DFs of the order patterns of length $L=4$ were obtained. The $\lambda $-DF
of the order pattern $[0,1,2,3]$ ($\#0$ for short) was calculated as the
mean value of the $\lambda $-DFs obtained from the various initial
conditions. This average value is compared to the exact $\lambda $-DF, $\phi
_{4}(\lambda )=\lambda ^{2}/(2-\lambda )$, in Fig.~\ref%
{figure:errorPDFestimation}, along with the corresponding standard
deviation.
\begin{figure}[tbp]
\centering
\includegraphics[scale=0.7]{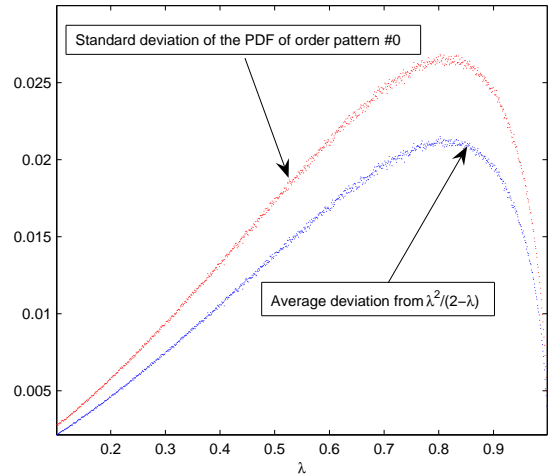}
\caption{Average deviation and standard deviation in the estimation of the
PDF of the order pattern $\#0$ for the skew tent map with respect to $%
\protect\lambda ^{2}/(2-\protect\lambda )$.}
\label{figure:errorPDFestimation}
\end{figure}

Fig.~\ref{figure:errorPDFestimation} spells out that, in the context of
finite precision computation, the perfect recovery of the control parameter
value using the $\lambda $-DF of order pattern $\#0$, is not feasible in
general if one can only resort to Gray codes. However, it is possible to
locate $\lambda $ up to an uncertainty interval. The width of this interval
can be upper bounded by the standard deviation of the $\lambda $-DF of the
order pattern $\#0$ since, according to Fig.~\ref{figure:errorPDFestimation}%
, it is bigger than the average error in the estimation of $\phi
_{4}(\lambda )$ for every value of $\lambda $ . Therefore, the estimation of
the control parameter comprises two stages:

\begin{enumerate}
\item An estimation of $\lambda $ is performed by dividing the given Gray
code, $\left\{ g_{i}\right\} _{i=0}^{M-1}$, $g_{i}\in \{0,1\}$, into a large
enough set of disjoint subsequences of length $N\gg 4$, say, $\left\{
g_{k\cdot N+i}\right\} _{i=0}^{N-1}$ for $k=0,1,\ldots ,K=\left\lfloor
M-N\right\rfloor -1$. For each such binary subsequence, a value of $\phi
_{4}(\lambda )$ is then computed as the relative frequency of the order
pattern $[0,1,2,3]$ using, of course, the Gray ordering (Sect.~\ref{sec:grayCodes}). Let $%
\bar{x}$ be the mean value of the resulting $\phi _{4}(\lambda )$'s. From $%
\phi _{4}(\lambda )=\lambda ^{2}/(2-\lambda )$, Eq.~%
\eqref{eq:orderPattern0skew}, it follows that the control parameter is
estimated as
\begin{equation}
\hat{\lambda}=\phi _{4}^{-1}(\bar{x})=\frac{-\bar{x}+\sqrt{\bar{x}^{2}+8\bar{%
x}}}{2}.
\end{equation}

\item If $\sigma $ is the standard deviation of the $\phi _{4}(\lambda )$
sampling, then
\begin{equation}
\lambda \in (\phi _{4}^{-1}(\bar{x}-\sigma ),\phi _{4}^{-1}(\bar{x}+\sigma
)).
\end{equation}
\end{enumerate}

\begin{figure}[tbh]
\includegraphics[scale=0.7]{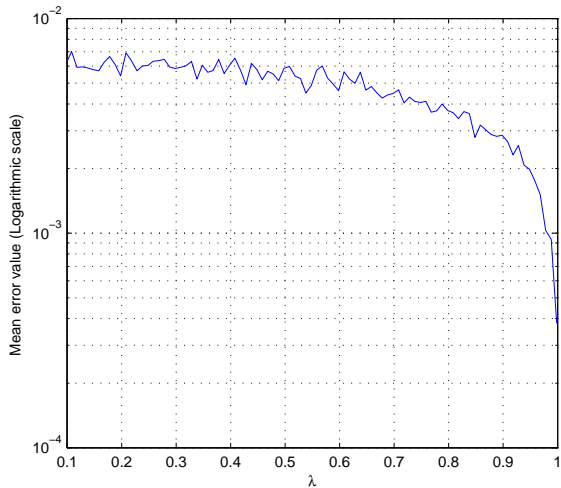}
\caption{Mean error value in the estimation of the control parameter of the
skew tent map.}
\label{figure:meanError}
\end{figure}
The specifics of this procedure refers to the skew tent map, but the
general strategy is the same, once a bijective $\lambda $-DF of an
order pattern is exactly known.

\begin{figure}[!htb]
\centering
\includegraphics[width=8cm,height=6cm]{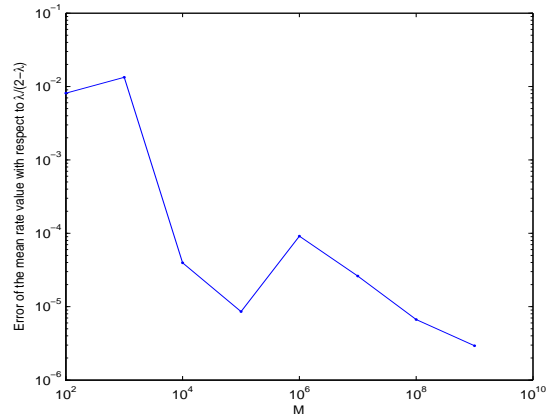}
\caption{Dependency of the error in the estimation of the rate of
occurrences of the order pattern $\#0$ with respect to the length of the
orbits.}
\label{figure:errorSymmetric}
\end{figure}

\begin{figure}[!htb]
\centering
\includegraphics[width=8cm,height=6cm]{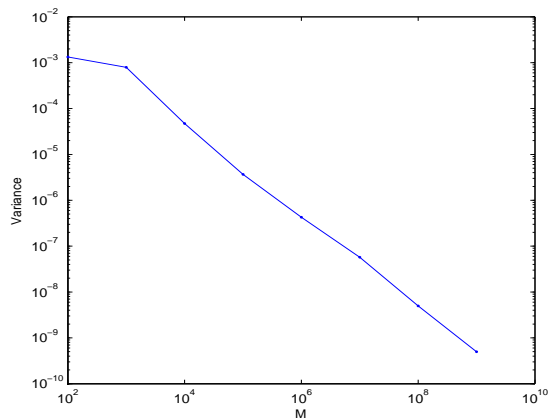}
\caption{Variance of the error in the estimation of the rate of occurrences
of the order pattern $\#0$ with respect to the length of the orbits.}
\label{figure:varianceSymmetric}
\end{figure}

In order to establish the accuracy of the procedure, some numerical
simulations with the skew tent map were done. For every value of the control
parameter, a group of $100$ different initial conditions were used in the
generation of the corresponding Gray codes. For each of these binary
sequences, the control parameter $\lambda $ was estimated as just explained.
The mean error of the estimation is shown in Fig.~\ref{figure:meanError}.
The average error lies always above $10^{-4}$, and can only be reduced by
the implementation of the procedure with extended-precision arithmetic
libraries. To prove this claim, the case of the symmetric tent map, i.e.,
the skew tent map for $\lambda =1/2$, will be now considered. For the
symmetric tent map the arithmetic is exact. Indeed, if $0.x_{0}x_{1}...x_{M}$%
, $x_{i}\in \{0,1\}$, is the expansion to base $2$ of $x\in \lbrack 0,1]$,
i.e.,
\begin{equation}
x=\sum_{n=0}^{M}\frac{x_{n}}{2^{n+1}},  \label{eq:xBin}
\end{equation}%
(numbers with finite binary expansions are called dyadic rationals), then
the action of the symmetric tent map amounts to a zero-bit dependent left
shift, to wit:
\begin{eqnarray}
\lefteqn{f_{1/2}(0.x_{0}x_{1}...x_{n}...x_{M}) =} \nonumber\\&& {}
=\left\{
\begin{array}{ll}
0.x_{1}x_{2}\ldots x_{n+1}\ldots x_{M-1}x_{M}, & \text{ if }x_{0}=0, \\
0.x_{1}^{\ast }x_{2}^{\ast }\ldots x_{n+1}^{\ast}\ldots x_{M-1}^{\ast}x_{M}, & \text{ if }x_{0}=1,%
\end{array}%
\right.
\end{eqnarray}%
where $x_{n}^{\ast }=1-x_{n}$. Therefore, if $x\in \lbrack 0,1]$ is
represented with $M$ bits and $x_{M}=1$, the orbit of $x$ collapses to $0$
after $M$ iterations of $f_{1/2}$, so $M$ can be considered the effective
length of the orbits to be used in an estimation of $\lambda =1/2$. For $%
L=3, $ the relative frequency of the order pattern $\#0$ ($[0,1,2]$ in this
case) was determined for a large set of random initial conditions $x$ and
increasing orbit lengths $M$. The convergence in average of this relative
frequency to $\phi _{3}(1/2)=1/3$ (see Eq. (\ref{eq:upperBound})) as $M$
increases, is confirmed by Fig.~\ref{figure:errorSymmetric}. At the same
time, the variance of the estimation steadily reduces with $M$, as shown in
Fig.~\ref{figure:varianceSymmetric}. In other words, a higher precision of
the arithmetic used in orbit generation and greater samples for the
subsequent control parameter estimation, clearly improves the results.

We conclude that the inaccuracies exposed above in our method to recover the
control parameter of maps of $\mathcal{F}_{2}$, based on the order patterns
of their coarse-grained orbits (specifically, in form of Gray codes), are
due to the shortcomings of finite precision arithmetic and finite
statistical sampling, but are not inherent to the method ---as proved with
the symmetric tent map.

\section{Conclusions}

\label{sec:conclusions} In this paper it was shown how to rebuild the $%
\lambda $-DFs of order patterns from Gray codes, the scope being the
estimation of the parameter $\lambda $. Gray codes are the 0-1 sequences
that result from the symbolic dynamic of unimodal maps with respect to the
left-right partition of the state space introduced by the critical point. We
have analyzed the $\lambda $-DFs of the order patterns of two unimodal
parametric maps: the logistic map (as representative of the subclass $%
\mathcal{F}_{1}$) and the skew tent map (as representative of the subclass $%
\mathcal{F}_{2}$). In the case of the logistic map, it turns out that this
technique can hardly deliver, on account of the complex and many-to-one
relation between $\lambda $ and those $\lambda $-DFs. On the contrary, this
relationship is simple, one-to-one, and analytically known for $\pi
=[0,1,...,L-1]$ in the case of the skew tent map. Our method improves
previous proposals for parameter estimation of unimodal maps in that a
knowledge of the critical point value is not needed. However, it demands
high computational precision and large amounts of data; in this regard, we
recommend the use of extended-precision libraries for good estimations. In
the ideal case of arbitrarily high precision, the estimated value of the
control parameter is arbitrarily close to the real one.

\section*{Acknowledgments}

The work described in this paper was supported by \textit{Minis\-terio de
Educaci\'on y Ciencia of Spain}, research grant SEG2004-02418, \textit{CDTI,
Minis\-terio de Industria, Turismo y Comercio of Spain} in collaboration
with Telef\'onica I+D, Project SEGUR@ with reference CENIT-2007 2004,
\textit{CDTI, Minis\-terio de Industria, Turismo y Comercio of Spain} in
collaboration with SAP, project HESPERIA (CENIT 2006-2009), and \textit{%
Ministerio de Ciencia e Innovaci\'on of Spain} in collaboration, project
CUCO (MTM2008-02194).

\bibliographystyle{chaos}

\end{document}